\documentclass[preprint,pre,aps,amsfonts,amsmath,showpacs,superscriptaddress,nofootinbib]{revtex4}
\pdfoutput=1
\usepackage{amsmath,amssymb,amscd}
\usepackage{mathrsfs}
\usepackage{graphicx}
\usepackage{bm}
\usepackage{color}

\newcommand{\e}{{\rm e}}
\newcommand{\E}{{\mathbb E}}

\newcommand{\bea}{\begin{eqnarray}}
\newcommand{\eea}{\end{eqnarray}}

\begin{document}

\title{Smoluchowski flux and Lamb-Lion Problems for Random Walks and L\'evy Flights with a Constant Drift}
\author{Satya N. Majumdar}
\email{satya.majumdar@lptms.u-psud.fr}
\affiliation{LPTMS, CNRS, Univ. Paris-Sud, Universit\'e Paris-Saclay, 91405 Orsay, France.}
\author{Philippe Mounaix}
\email{philippe.mounaix@cpht.polytechnique.fr}
\affiliation{CPHT, CNRS, Ecole
Polytechnique, IP Paris, F-91128 Palaiseau, France.}
\author{Gr\'egory Schehr}
\email{gregory.schehr@lptms.u-psud.fr}
\affiliation{LPTMS, CNRS, Univ. Paris-Sud, Universit\'e Paris-Saclay, 91405 Orsay, France.}
\date{\today}
\begin{abstract}
{We consider non-interacting particles (or lions) performing one-dimensional random walks or L\'evy flights (with L\'evy index $1 < \mu \leq 2$) in the presence of a constant drift $c$. Initially these random walkers are uniformly distributed over the positive real line $z\geq 0$ with a density $\rho_0$. At the origin $z=0$ there is an immobile absorbing trap (or a lamb), such that when a particle crosses the origin, it gets absorbed there. Our main focus is on (i) the flux of particles $\Phi_c(n)$ out of the system (the ``Smoluchowski problem'') and (ii) the survival probability $S_c(n)$ of the trap or lamb (the ``lamb-lion problem'') until step $n$. We show that both observables can be expressed in terms of the average maximum $\mathbb{E}[M_c(n)]$ of a single random walk or L\'evy flight after $n$ steps. This allows us to obtain the precise asymptotic behavior of both $\Phi_c(n)$ and $S_c(n)$ analytically for large $n$ in the two problems, for any value of $1<\mu\leq 2$ and $c \in {\mathbb{R}}$. In particular, for $c>0$, we show the rather counterintuitive result that for $1< \mu < 2$, $S_{c>0}(n \to \infty)$ vanishes as $S_{c>0}(n \to \infty) \approx \exp\left(-\lambda \, n^{2-\mu}\right)$, where $\lambda$ is a $\mu$-dependent positive constant, while for standard random walks (i.e., with $\mu = 2$), $S_{c>0}(n \to \infty) \to K_{RW} > 0$, as expected. Our analytical results are confirmed by numerical simulations.}
\end{abstract}
\maketitle
%
%
\section{Introduction}\label{sec1}
Brownian motion and random walks have been fundamental cornerstones of statistical physics for more than a century. While Brownian motion is the building block for processes with continuous trajectories, random walks underlie processes with countable increments built from jumps occurring at discrete time steps. Both have an outstanding number of applications in a large range of fields \cite{chandra,feller,satya_reviewBM,duplantier_review,Yor_book}. For instance, adding a drift to the standard Brownian motion and random walks allows to study the effect of linear trends in several fluctuating observables of interest such as statistics of records \cite{revue_record, record_drift} with specific applications to finance~\cite{satya_jp,WBK11,record_multiple}. In this paper, we study the effect of a constant drift in two random walk problems, namely the celebrated Smoluchowski flux problem and the so called ``lamb-lion'' problem that is defined precisely later. These two problems have been studied before for Brownian motions without drift and several exact results are known in this case. In this paper we generalize these results to (i) Brownian motions in the presence of a constant drift $c$ and also (ii) to the case  
where the walkers perform random walks/L\'evy flights in the presence of a constant drift $c$.  

Consider first a single Brownian motion in the present of a constant drift $c$ in one dimension. Its position $x(t)$ on the line evolves, with time $t$, according to
\begin{equation}\label{eqBM}
\frac{dx(t)}{dt}= c + \sqrt{2D}\, \eta(t) \;,
\end{equation}
starting from $x(0)=0$, where the drift $c$ is a fixed number, $D$ is the diffusion constant, and $\eta(t)$ is a Gaussian white noise with zero mean and delta correlator, $\langle \eta(t)\eta(t')\rangle = \delta(t-t')$. For a discrete-time random walk (RW) with a constant drift $c$, the counterpart of Eq.\ (\ref{eqBM}) reads
\begin{equation}\label{eqRW}
x_i =x_{i-1}+c+\eta_i \;,
\end{equation}
giving the evolution of the position $x_i$ of the walker after $i$ steps, for $i\ge 1$, starting at $x_0 =0$. The jump increments $\eta_i$'s are independent and identically distributed (i.i.d.) random variables, each drawn from a symmetric and piecewise continuous probability distribution function (PDF) $f(\eta)$. Let $M_c(t)\ge 0$ denote the maximum position of the walker up to time $t$, where $t\in\mathbb{R}^+$ if the walker position is given by Eq.\ (\ref{eqBM}) (continuous-time Brownian motion), or $t\in\mathbb{N}$ if it is given by Eq.\ (\ref{eqRW}) (discrete-time RW). The main motivation of this work is to show how results about the expected maximum, $\E[M_c(t)]$, can be used to solve other, {\it a priori} unrelated, problems. To this end, we will take advantage of our recent exhaustive study of the asymptotic large time behavior of $\E[M_c(t)]$\ \cite{MMS2018}. The two problems we solve in this paper follow from the situation where one considers a semi-infinite line $z\in[0,\infty)$ with a single immobile trap, or absorber, at the origin $z=0$. Initially, the full semi-infinite line to the right of the trap is filled uniformly by non-interacting particles with uniform density $\rho_0$. Each particle performs an independent random walk as in Eqs.\ (\ref{eqBM}) or\ (\ref{eqRW}) with a drift~$c$. When a particle crosses the trap at the origin, it gets absorbed there. There are then two natural and interesting questions related to this physical situation.

\begin{figure}
\includegraphics[width = 0.7\linewidth]{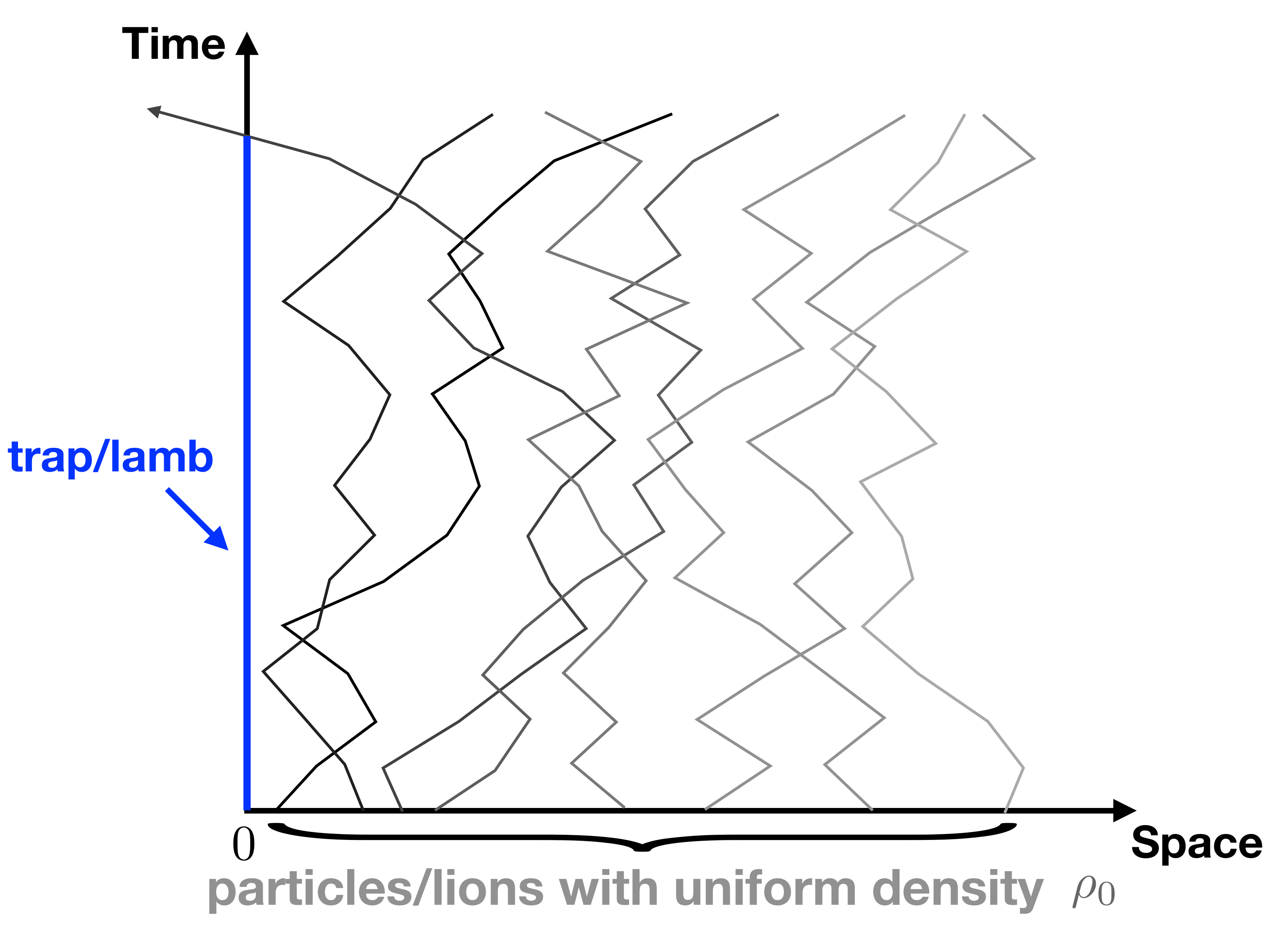}
\caption{Illustration of the Smoluchowski and lamb-lion problems. A trap or a lamb is immobile at the origin at $z=0$ while the positive axis is filled with non-interacting particles or lions (with an initial uniform density $\rho_0$) that perform a discrete random walk with a drift $c$ as in Eq. (\ref{eqRW}). The particle 
gets absorbed when it crosses $z=0$. Our main focus is on (i) the flux of particles through the trap at the origin, $\Phi_c(n)$, after $n$ steps (the so called ``Smoluchowski'' problem) and (ii) the survival probability $S_c(n)$ of the trap or lamb, i.e. the probability that no particle has hit the origin  
up to step $n$ (the so-called ``lamb-lion'' problem). The blue line segment on the time axis represents the lifetime of the trap or lamb.}\label{Fig_lamb_lion}
\end{figure}

The first one is the one-dimensional version of the celebrated Smoluchowski problem\ \cite{Smoluchowski} where one is interested in the net flux $\Phi_c(t)$ of particles out of the system up to time $t$. Interestingly, on can show that this flux is directly related to the expected maximum via the relation
\begin{equation}\label{eqFlux}
\Phi_c(t)=\rho_0\, \E[M_c(t)] \;,
\end{equation}
the derivation of which is recalled below in Sec.\ \ref{sec3.1} (for continuous-time Brownian motion) and in Sec.\ \ref{sec3.2} (for discrete-time RW).

The second question, in the same physical setting, is the survival probability of the trap, $S_c(t)$, up to time $t$. More precisely, $S_c(t)$ is the probability that none of the particles, initially uniformly distributed with density $\rho_0$, has hit the origin up to time $t$. For the case $c=0$, the survival probability $S_0(t)$ has been computed exactly and is generally known as the target-annihilation problem \cite{BZK84,Tac83,BO87,FM2012} (for a recent review see \cite{pers_review}). Sometimes, this problem is also known as the lamb-lion problem: an immobile lamb is located at the origin $z=0$, while $N$ lions with initial positions uniformly distributed over the interval $z\in[0,L]$ undergo independent Brownian motions. A variant of this problem where the lamb itself performs Brownian motion has generated a considerable interest and still remains unsolved to a large extent~\cite{Redner,BL88,BB2002a,BB2002b,OBCM2002,BB2003,BMB2003,MB2003}.  
Here, we consider this target annihilation problem with an immobile lamb at the origin while the lions perform (i) independent Brownian motions with a constant drift $c$ and (ii)  independent random walks/L\'evy flights with a constant drift $c$. Our main objective is to compute the survival probability $S_c(t)$ of the lamb in these two cases in the thermodynamic limit: $L,\, N\to \infty$ keeping $\rho_0 =N/L$ fixed.  
In this thermodynamic limit, we show below that $S_c(t)$ is related to the expected maximum $\E[M_c(t)]$ of a single random walk/L\'evy flight that starts at the origin (see Secs.\ \ref{sec4.1} and\ \ref{sec4.2})
\begin{equation}\label{eqSurviv}
S_c(t)=\exp\left[ -\Phi_{-c}(t)\right]
=\exp\left[ -\rho_0\, \E[M_{-c}(t)]\right] \;,
\end{equation}
where we have used Eq.\ (\ref{eqFlux}) in the last equality. For the case $c=0$, this relation connecting the survival probability to the expected maximum $S_0(t) = \exp[-\rho_0\E(M_0(t))]$ was established in Ref. \cite{FM2012}. Note that our result in Eq. (\ref{eqSurviv}) is valid both for biased Brownian motion in continuous time as well as biased random walks/L\'evy flights -- in the latter case $t$ refers to the discrete time step $n$.

The relations in Eqs.\ (\ref{eqFlux}) and\ (\ref{eqSurviv}) establish a nice link between the expected maximum $\E[M_c(t)]$, the integrated flux $\Phi_c(t)$, and the survival probability $S_c(t)$, providing thus two nontrivial physical applications for the expected maximum of a random walk with a drift. These questions are quite well understood for the Brownian motion\ (\ref{eqBM}) which, by virtue of the central limit theorem, describes the large $n$ limit of random walks\ (\ref{eqRW}) with jumps having a well defined second moment, $\int_{-\infty}^{+\infty}\eta^2 f(\eta)\, d\eta < +\infty$. In the absence of a drift, i.e. for $c=0$, it is well known that the expected maximum behaves as\ \cite{Yor_book}
\begin{equation}\label{eqBMEM0}
\E[M_0(t)]=\frac{2}{\sqrt{\pi}}\sqrt{Dt}.
\end{equation}
For $c\ne 0$, the expected maximum behaves quite differently. Indeed, for large $t$, one has (see\ \cite{MMS2018} for a simple derivation of both\ (\ref{eqBMEM0}) and\ (\ref{eqBMEMc}))
\begin{equation}\label{eqBMEMc}
\E[M_c(t)]\sim\theta(c)\, ct +\frac{D}{\vert c\vert}\ \ \ \ \ (t\to\infty),
\end{equation}
where $\theta(c)$ is the Heaviside step function, $\theta(c)=1$ if $c>0$ and $\theta(c)=0$ if $c<0$. From the results in Eqs.\ (\ref{eqBMEM0}) and\ (\ref{eqBMEMc}), together with the relation in Eq.\ (\ref{eqSurviv}), it follows that the large time behavior of the survival probability $S_c(t)$ in the case of Brownian motion is given by
\begin{equation}\label{eqBMSurviv}
S_c(t)\sim\left\lbrace
\begin{array}{ll}
K_{BM}\exp\left(-\rho_0 \vert c\vert t\right),&c<0 \\
\exp\left(-\frac{2\rho_0}{\sqrt{\pi}}\sqrt{Dt}\right),&c=0 \\
K_{BM},&c>0
\end{array}\right.\ \ \ \ \ (t\to\infty),
\end{equation}
where $K_{BM}=\exp(-D\rho_0/\vert c\vert)$ is a constant, which shows in particular that if $c>0$ then the lamb (or trap) will survive with a finite probability as $t\to\infty$.

In our previous paper\ \cite{MMS2018} we addressed the interesting question of how Eqs.\ (\ref{eqBMEM0}) and\ (\ref{eqBMEMc}) get affected when the continuous-time Brownian motion\ (\ref{eqBM}) is replaced by a discrete-time Markov process like in\ (\ref{eqRW}), including L\'evy flights. To this end we considered jump PDFs, $f(\eta)$, the Fourier transform of which $\hat f(k) = \int_{-\infty}^{+\infty} e^{ik\,\eta}\,f(\eta) \, d\eta$ has the small $k$ behavior
\begin{equation}\label{eq1.3}
\hat f(k) = 1 - |a k|^\mu + O\left(|k|^{\nu}\right),
\end{equation}
where $a>0$ is the characteristic length scale of the jumps, $1 < \mu \leq 2$ is the L\'evy index, and the subleading exponent $\nu> \mu$, (note that we need $\mu >1$ for $\E[M_c(t)]$ to exist). 
For $\mu = 2$, the variance of the jump distribution $\sigma^2 = \int_{-\infty}^{+\infty} \eta^2 f(\eta) d\eta$ is finite and $a = \sigma /\sqrt{2}$. In this case, the suitably scaled RW converges to a Brownian motion as $t\rightarrow +\infty$. On the other hand, for $0 < \mu < 2$, $f(\eta)$ is a fat-tailed distribution, $f(\eta) \propto |\eta|^{-1-\mu}$ ($\eta \to \infty$), and the RW\ (\ref{eqRW}) is a L\'evy flight of index $\mu$. In the following we write $n\equiv t\in\mathbb{N}$ the (discrete) time variable pertaining to the discrete-time RW\ (\ref{eqRW}). In the absence of a drift, i.e. for $c=0$, it was found in\ \cite{AS2005} that the discrete-time counterpart of Eq.\ (\ref{eqBMEM0}) reads
\begin{equation}\label{eqRWEM0}
\E [M_0(n)]\sim\frac{a\mu\Gamma(1-1/\mu)}{\pi} n^{1/\mu}+a\gamma\ \ \ \ \ (n\rightarrow +\infty),
\end{equation}
with
\begin{equation}\label{gamma}
\gamma=\frac{1}{\pi}\int_0^{+\infty}\ln\left(\frac{1-\hat{f}(q/a)}{q^\mu}\right)\, \frac{dq}{q^2}.
\end{equation}
It is important to notice that, while the leading term on the right-hand side of Eq.\ (\ref{eqRWEM0}) gives the leading large $n$ behavior of $\E[M_0(n)]$ correctly for all $\nu >\mu$ (see Eq.\ (\ref{eq1.3})), the next subleading correction needs $\nu>\mu +1$ to be a constant. In the complementary domain $\mu <\nu <\mu +1$, one finds that $\gamma$ in Eq.\ (\ref{eqRWEM0}) must be replaced with a term growing as $n^{1-(\nu -1)/\mu}$\ \cite{GLM2017}. For simplicity, in the following we will always assume that in the absence of a drift, $c=0$, the inequality $\nu >\mu +1$ is fulfilled. (Otherwise we assume the less stringent condition $\nu\ge 2$). Note also that for $\mu =2$, the first term on the right-hand side of\ (\ref{eqRWEM0}) coincides with Eq.\ (\ref{eqBMEM0}) (with $D=a^2$ and $t=n$). On the other hand, for $c\ne 0$ the large $n$ behavior of $\E[M_c(n)]$ depends on the value of the L\'evy index $\mu$ (see\ \cite{MMS2018} for details). For $\mu =2$ one has
\begin{equation}\label{eqRWEMc2}
\E[M_c(n)]\sim\theta(c)\, cn + \vert c\vert\,\kappa_c\ \ \ \ \ (n \to \infty) \;,
\end{equation}
with
\begin{equation}\label{kappac}
\kappa_c =\frac{1}{2\pi}\frac{\partial}{\partial\lambda}\left.\int_{-\infty}^{+\infty}
\frac{\ln\lbrack 1-\hat{f}(q/c){\rm e}^{-iq}\rbrack}{\lambda +iq}\, dq\right\vert_{\lambda =0} \;,
\end{equation}
and for $1<\mu<2$, one finds
\begin{equation}\label{eqRWEMcMu}
\E[M_c(n)]\sim\theta(c)\, cn +  \vert c\vert\frac{C \,n^{2-\mu}}{2-\mu}\ \ \ \ \ (n\to\infty) \;,
\end{equation}
where 
\begin{equation}\label{const1}
C=\frac{\Gamma(\mu -1)}{\pi}\sin\left(\frac{\pi\mu}{2}\right)\left(\frac{a}{\vert c\vert}\right)^\mu \;.
\end{equation}
Using then Eqs.\ (\ref{eqRWEM0}),\ (\ref{eqRWEMc2}), and\ (\ref{eqRWEMcMu}) on the right-hand side of the general relation\ (\ref{eqSurviv}) yields the large $n$ behavior of the survival probability $S_c(n)$ in the case of a discrete-time lion RW. For a random walk with $\mu =2$, one gets
\begin{equation}\label{eqRWSurviv2}
S_c(n)\sim\left\lbrace
\begin{array}{ll}
K_{RW}\exp\left(-\rho_0 \vert c\vert n\right),&c<0 \\
K_{RW}^{(0)}\exp\left(-\frac{2a\rho_0}{\sqrt{\pi}}\sqrt{n}\right),&c=0 \\
K_{RW},&c>0
\end{array}\right.\ \ \ \ \ (n\to\infty),
\end{equation}
where $K_{RW}^{(0)}=\exp(-a\rho_0\gamma)$ and $K_{RW}=\exp(-\vert c\vert\rho_0\kappa_c)$ are constants (note that $\kappa_{-c}=\kappa_c$). On the other hand, for a L\'evy flight with $1<\mu <2$, one finds
\begin{equation}\label{eqRWSurvivMu}
S_c(n)\sim\left\lbrace
\begin{array}{ll}
\mathscr{S}_{<}(n)\exp\left(-\rho_0 \vert c\vert n\right),&c<0 \\
K_{RW}^{(0)}\exp\left(-\frac{a\rho_0\mu\Gamma(1-1/\mu)}{\pi}n^{1/\mu}\right),&c=0 \\
\mathscr{S}_{>}(n)\exp\left(-\frac{c\rho_0 C}{2-\mu}n^{2-\mu}\right),&c>0
\end{array}\right.\ \ \ \ \ (n\to\infty),
\end{equation}
where $\mathscr{S}_{<}(n)$ and $\mathscr{S}_{>}(n)$ are slowly varying compared respectively with the exponential of $n$ and $n^{2-\mu}$, with $\mathscr{S}_{<}(n)=\mathscr{S}_{>}(n)\exp[-n^{2-\mu}\vert c\vert\rho_0C/(2-\mu)]$. Note that the numerical value of the constant $K_{RW}^{(0)}$ in Eq.\ (\ref{eqRWSurvivMu}) is not the same as in Eq.\ (\ref{eqRWSurviv2}), as $\gamma$ in Eq.\ (\ref{gamma}) depends on $\mu$. The growth of $\E[M_c(n)]$ in Eq.\ (\ref{eqRWEMcMu}) for $1 < \mu < 2 $ and $c<0$ is somewhat unexpected since, in this case, the process $y_n = x_n - c \,n $, with $x_n$ given by Eq. (\ref{eqRW}), converges to a symmetric L\'evy flight for large $n$ with $y_n = O(n^{1/\mu})$, typically. This is much smaller than the drift term $c \,n$ and one could thus legitimately expect $\E[M_c(n)]$ to approach a constant for large $n$, like in the $\mu=2$ case in Eq.\ (\ref{eqRWEMc2}). This is not the case: for $c<0$, although the walker will typically drift to $-\infty$, she/he will always perform rare big jumps that will contribute to higher and higher values of $\E[M_c(n)]$ significantly as $n$ increases. An interesting consequence of this result on the lamb-lion problem, for lions undergoing independent L\'evy flights with a constant drift, is that the survival probability of the lamb (or trap) still decays to zero in the presence of a positive drift, as can be seen in the third Eq.\ (\ref{eqRWSurvivMu}). Albeit a bit counterintuitive, this result is now easy to understand: lions will always perform rare big jumps that will overcompensate for their linear drift away from the lamb.

Note that the large $n$ behavior of $\E[M_c(n)]$ in Eq.\ (\ref{eqRWEMcMu}), as well as the corresponding expressions of $\Phi_c(n)$ and $S_c(n)$, via Eqs.\ (\ref{eqFlux}) and\ (\ref{eqSurviv}), are leading behaviors only. As we will see in the following, the subleading corrections to $\Phi_c(n)$ and $S_c(n)$ (i.e. the functions $\mathscr{S}_{\lessgtr}(n)$ is Eq.\ (\ref{eqRWSurvivMu})) can also be obtained from the surviving subleading terms of $\E[M_c(n)]$ given in Ref.\ \cite{MMS2018} [see also Eqs. (\ref{prefactor1})-(\ref{prefactor3}) below].

The outline of the paper is as follows. In section\ \ref{sec2} we recall some useful known results about the statistics of the maximum $M_c(t)$. The case of a $1D$ continuous-time Brownian motion with a constant drift $c$ is considered in section\ \ref{sec2.1}. Section\ \ref{sec2.2} is devoted to $1D$ discrete-time random walks ($t=n\in\mathbb{N}$) with a constant drift $c$. The results of Ref.\ \cite{MMS2018} giving the large $n$ asymptotic expansion of $\E[M_c(n)]$ are recalled, without demonstration, including all the terms surviving the large $n$ limit, for both random walks with $\mu =2$ and L\'evy flights with $1<\mu <2$. The relations given in equations\ (\ref{eqFlux}) and\ (\ref{eqSurviv}) are derived in sections\ \ref{sec3} and\ \ref{sec4}, respectively, first for a Brownian motion (Secs.\ \ref{sec3.1} and\ \ref{sec4.1}), then in the case of random walks and L\'evy flights (Secs.\ \ref{sec3.2} and\ \ref{sec4.2}). In both cases, the large time asymptotic behavior of $\Phi_c(t)$ and $S_c(t)$ is obtained from the one of $\E[M_c(t)]$ as a function of the drift $c$. In section\ \ref{sec4.3}, we discuss and solve an apparent paradox about the first equality\ (\ref{eqSurviv}) linking the survival probability of the lamb in the lamb-lion problem to the total flux out of the system in the Smoluchowski problem. In section\ \ref{sec5}, we verify our analytical predictions via numerical simulations. Finally, we conclude in section\ \ref{sec6}.
%
%
\section{Reminder of some useful results}\label{sec2}
In this section we briefly recall some useful results about the statistics of the maximum of a $1D$ continuous-time Brownian motion and of $1D$ discrete-time random walks (and L\'evy flights), both in the presence of a constant drift $c$.
%
%
\subsection{$\bm{1D}$ Brownian motion}\label{sec2.1}
First, we consider the case of a $1D$ Brownian motion. The interested reader is referred to the section\ \ref{sec2} of Ref.\ \cite{MMS2018} for details. Let $x(t)$ be a biased Brownian motion on a line starting from $x=0$ at $t=0$ and evolving according to Eq.\ (\ref{eqBM}). Write $M_c(t)=\max_{0\le\tau\le t}\lbrace x(\tau)\rbrace$ the maximum of this process up to time $t$, where the subscript $c$ stands for the presence of the constant drift $c$ in Eq.\ (\ref{eqBM}). Let $Q_c(z,t)\equiv {\rm Prob}[M_c(t)\le z]$ denote the cumulative distribution of $M_c(t)$. Clearly, $Q_c(z,t)$ is also the probability that the Brownian trajectory stays below $z\ge 0$ up to time $t$. Define $y(t)= z-x(t)$ a Brownian motion with a drift $-c$ starting from $y=z$ at $t=0$. Hence, $Q_c(z,t)$ is also the probability that the process $y(t)$ (with drift $-c$) stays positive (does not cross zero) up to time $t$. It is then easy to write a backward Fokker-Planck evolution for $Q_c(z,t)$\ \cite{pers_review},
\begin{equation}
\frac{\partial Q_c(z,t)}{\partial t}= D\, \frac{\partial^2 Q_c(z,t)}{\partial 
z^2} - c\, \frac{\partial Q_c(z,t)}{\partial z}  \, ,
\label{bfp_max.1}
\end{equation}
valid for $z\ge 0$ with the boundary conditions
\begin{equation}
Q_c(z=0,\, t)=0\, ; \quad {\rm and} \quad Q_c(z\to \infty,\, t)=1
\label{bc.1}
\end{equation}
together with the initial condition
\begin{equation}
Q_c(z,\,  t=0)=1 \quad {\rm for}\,\, z>0 \, .
\label{ic.1}
\end{equation}
This linear equation can be solved exactly. One finds,
\begin{equation}
Q_c(z,t)= \frac{1}{2}\left[{\rm erfc}\left(- 
\frac{z-ct}{\sqrt{4Dt}}\right)- 
e^{c\,z/D}\, {\rm erfc}\left(\frac{z+ct}{\sqrt{4Dt}}\right)\right]\, ,
\label{sol_max.1}
\end{equation}
where ${\rm erfc}(x)= \frac{2}{\sqrt{\pi}}\, \int_x^{+\infty} e^{-u^2}\, du$ is the complementary error function. It is easy to see from Eq. (\ref{sol_max.1}) that for $c>0$, the cumulative distribution $Q_{c>0}(z,t)$ is always time-dependent, while for $c<0$, it approaches a time-independent stationary distribution
\begin{equation}
Q_{c<0}(z,t) \xrightarrow[t\to \infty]{} 1- \exp\left[- \frac{|c|\, 
z}{D}\right]\,.
\label{stat_dist.1}
\end{equation}
The PDF of $M_c(t)$ is the derivative $\partial_z Q_c(z,t)$, hence the expected maximum is given by $\E[M_c(t)]= \int_0^{+\infty} z\, \partial_z Q_c(z,t)\, dz$, which, via integration by parts, can be written as
\begin{equation}
\E[M_c(t)] = \int_0^{+\infty} \left[1- Q_c(z,t)\right]\, dz \,.
\label{mean_max.1}
\end{equation}
From Eq.\ (\ref{mean_max.1}) and the result in Eq. (\ref{sol_max.1}), it is possible to compute the expected maximum $\E[M_c(t)]$. One gets an exact expression valid for all $c$ and all $t$, (see\ \cite{MMS2018} for details)
\begin{equation}
\E[M_c(t)] = \theta(c)\, ct + \sqrt{D\, t}\, {\cal G}_2\left(\frac{|c|\sqrt{t}}{\sqrt{D}}\right),
\label{mean_max_scaling.1}
\end{equation}
where the scaling function ${\cal G}_2(u)$ is exactly given by
\begin{eqnarray}\label{mean_max_scaling.2}
{\cal G}_2(u)&=&
\frac{4}{u\sqrt{\pi}}\, \int_0^{u/2} dv\, \left[e^{-v^2}- \sqrt{\pi}\, v\, {\rm erfc}(v)\right] \nonumber \\
&=&\frac{1}{\sqrt{\pi}}\, e^{-u^2/4} + \frac{1}{u}\, {\rm erf}\left(\frac{u}{2}\right)- 
\frac{u}{2}\, {\rm erfc}\left(\frac{u}{2}\right)\, ,
\end{eqnarray}
with the asymptotics ${\cal G}_2(u) \to 2/\sqrt{\pi}$ as $u\to 0$ and ${\cal G}_2(u) \to 1/u$ as $u \to \infty$. For $c=0$ (no drift), Eqs.\ (\ref{mean_max_scaling.1}) and\ (\ref{mean_max_scaling.2}) yield simply
\begin{equation}\label{eqBMEM0bis}
\E[M_0(t)]=\frac{2}{\sqrt{\pi}}\sqrt{Dt},
\end{equation}
while for $c\ne 0$ and large $t$, one finds
\begin{equation}\label{eqBMEMcbis}
\E[M_c(t)]\sim\theta(c)\, ct +\frac{D}{\vert c\vert}\ \ \ \ \ (t\to\infty).
\end{equation}
Thus, for a positive drift $c>0$, the expected maximum increases linearly with $t$ (with speed $c$), while for a negative drift $c<0$, it approaches a constant $D/|c|$ as $t\to \infty$. The equations~(\ref{eqBMEM0bis}) and\ (\ref{eqBMEMcbis}) coincide respectively with Eqs.\ (\ref{eqBMEM0}) and\ (\ref{eqBMEMc}) in the introduction.
%
%
\subsection{$\bm{1D}$ Random walks and L\'evy flights}\label{sec2.2}
Now, we consider the case of a $1D$ discrete-time random walk (including L\'evy flights). The position $x_n$ of the walker at step $n$ evolves in discrete time step by the Markov rule\ (\ref{eqRW}), starting initially at $x_0 =0$. Write $M_c(n)=\max_{0\le m\le n}\lbrace x_m \rbrace$ the maximum of this process up to step $n$, with $n,\, m\in\mathbb{N}$ and where the subscript $c$ stands for the presence of the constant drift $c$ in Eq.\ (\ref{eqRW}). Like in the case of the Brownian motion, in the following we will need the evolution equation for the cumulative distribution $Q_c(z,n)\equiv {\rm Prob}[M_c(n)\le z]$. To this end, we use the fact that $Q_c(z,n)$ denotes the probability that the walker does not cross the level $z$ up to step $n$ and we define the process $y_n= z-x_n$ with $y_0 =z$. Thus, $Q_c(z,n)$ is also the probability that the process $y_n$, starting at $y_0 =z$ and in the presence of a drift $-c$, stays positive (does not cross zero) up to step $n$. Given the evolution\ (\ref{eqRW}), it follows that $y_n$ evolves via $y_n=y_{n-1}-c-\eta_n$ and using the fact that $\eta_n$ and $-\eta_n$ have the same PDF $f(\eta)$ (due to the symmetric nature of the noise), one finds that $Q_c(z,n)$ satisfies the recursion relation
\begin{equation}\label{binteq_max}
Q_c(z,n)=\int_0^{+\infty}Q_c(z^\prime ,n-1)\, f(z^\prime -z+c)\, dz^\prime ,
\end{equation}
with the initial condition $Q_c(z,0)=1$ for all $z\ge 0$. The integral equation\ (\ref{binteq_max}) is the discrete-time counterpart of the backward Fokker-Planck equation\ (\ref{bfp_max.1}) for the Brownian motion. The relation between $Q_c(z,n)$ and the expected maximum $\E[M_c(n)]$ is of course the same as in the continuous-time setting [see Eq. (\ref{mean_max.1})],
\begin{equation}
\E[M_c(n)] = \int_0^{+\infty} \left[1- Q_c(z,n)\right]\, dz \;.
\label{mean_max.2}
\end{equation}
The solution of the integral equation\ (\ref{binteq_max}) for arbitrary $f(\eta)$ is far from explicit and we will not use it to get $\E[M_c(n)]$ from Eq.\ (\ref{mean_max.2}), unlike the line followed in the Brownian motion case. Nevertheless, as we will see in the next two sections, Eqs.\ (\ref{binteq_max}) and\ (\ref{mean_max.2}) are still crucial to derive the Eqs.\ (\ref{eqFlux}) and\ (\ref{eqSurviv}) relating $\E[M_c(n)]$ to $\Phi_c(n)$ and $S_c(n)$.

Extracting the large $n$ behavior of $\E[M_c(n)]$ is a highly technical, non trivial, task that was carried out in Refs.\ \cite{MMS2018,AS2005} by using a suitably generalized version of the Pollaczek-Spitzer formula, instead of trying to solve Eq.\ (\ref{binteq_max}) directly. Here, we just recall the results without demonstration. The interested reader is referred to Refs.\ \cite{MMS2018,AS2005} for details.

In the absence of a drift, $c=0$, Comtet and Majumdar found that for $1<\mu\le 2$ and $\nu >\mu +1$ [see the remark below Eq.\ (\ref{gamma})], the large $n$ behavior of $\E[M_0(n)]$ is given by\ \cite{AS2005}
\begin{equation}\label{RWasym0}
\E [M_0(n)]=\frac{a\mu\Gamma(1-1/\mu)}{\pi} n^{1/\mu}+a\gamma
+O\left(\frac{1}{n^{1-1/\mu}}\right)\ \ \ \ \ (n\rightarrow +\infty),
\end{equation}
with $\gamma$ given in Eq.\ (\ref{gamma}). In the presence of a drift, $c\ne 0$, the large $n$ behavior of $\E[M_c(n)]$ depends on the value of the L\'evy index $\mu$. For $1<\mu <2$, with $\mu\ne 1+1/p$ for any integer $p$, one finds\ \cite{MMS2018}
\begin{equation}\label{RWasymMu.1}
\E[M_c(n)]=\theta(c)\, cn+\vert c\vert\sum_{m=1}^{[1/(\mu -1)]}
\frac{C_m n^{1-m(\mu -1)}}{1-m(\mu -1)} +\vert c\vert\kappa_c
+O\left(\frac{1}{n^{\mu-1}}\right)\ \ \ \ \ (n\rightarrow +\infty),
\end{equation}
where $[1/(\mu -1)]$ denotes the integer part of $1/(\mu -1)$, with
\begin{equation}\label{constCm}
C_m=(-1)^{m+1}\sin\left(\frac{m\pi\mu}{2}\right)\, \frac{\Gamma(m\mu-1)}{\pi m!}
\, \left(\frac{a}{\vert c\vert}\right)^{m\mu},
\end{equation}
and where $\kappa_c$ is a computable constant (see Section 5.2 in Ref.\ \cite{MMS2018} for details. Note that $\kappa_c$ in Eq.\ (\ref{RWasymMu.1}) corresponds to $\kappa_c +\Delta\kappa_c$ in\ \cite{MMS2018}). For $1<\mu <2$, with $\mu =1+1/p$ for some integer $p$, one finds that the equation\ (\ref{RWasymMu.1}) must be replaced by \cite{MMS2018}
\begin{equation}\label{RWasymMu.2}
\E[M_c(n)]=\theta(c)\, cn+\vert c\vert\sum_{m=1}^{p-1}
\frac{p\, C_m n^{1-m/p}}{p-m} +\vert c\vert C_p\ln n+\vert c\vert\kappa_c
+O\left(\frac{1}{n^{1/p}}\right)\ \ \ \ \ (n\rightarrow +\infty).
\end{equation}
Note that the leading terms in the equations\ (\ref{RWasymMu.1}) and\ (\ref{RWasymMu.2}) coincide with the asymptotics\ (\ref{eqRWEMcMu}), as it should be. Finally, for $\mu =2$ one has\ \cite{MMS2018}
\begin{equation}\label{RWasymMu=2}
\E[M_c(n)]=\theta(c)\, cn+\vert c\vert\kappa_c
+O\left(\frac{h(n)}{n^2}{\rm e}^{-nI(\vert c\vert)}\right)\ \ \ \ \ (n\rightarrow +\infty),
\end{equation}
with $\kappa_c$ given in Eq.\ (\ref{kappac}), $I(\vert c\vert)>0$, and $h(n)$ is subdominant with respect to the decaying exponential. Both $I(x)$ and $h(n)$ depends on the jump distribution $f(\eta)$ and there is no generic expressions for these two functions. The large $n$ behavior\ (\ref{RWasymMu=2}) coincides with the asymptotics\ (\ref{eqRWEMc2}).

From the results in Eq.\ (\ref{eqRWEM0}) (for $c=0$) and Eqs.\ (\ref{eqRWEMc2}, \ref{eqRWEMcMu}) (for $c\ne 0$) it is clear that the two limits $n\to\infty$ and $c\to 0$ do not commute, hence the small $c$ limit of the large $n$ behavior of $\E[M_c(n)]$ is singular. This suggests the existence of a scaling regime describing the crossover between the leading large $n$ behaviors of $\E[M_c(n)]$ for $c=0$ and a small nonzero $c$. The corresponding scaling form has been determined in Sec. 5.4 of Ref.\ \cite{MMS2018}. One finds
\begin{equation}\label{cross_3}
\E[M_c(n)]\sim\theta(c)\, cn +an^{1/\mu} {\cal G}_\mu\left(\frac{|c|}{a} n^{1-1/\mu} \right) \:,
\end{equation}
with
\begin{equation}\label{cross_4}
{\cal G}_\mu(u) = \frac{\mu}{\mu-1} u^{-\frac{1}{\mu-1}} \int_0^u dy \, y^{\frac{2-\mu}{\mu-1}} \int_0^{+\infty} dx \, x\, \, f_{S,\mu}(x+y) \;,
\end{equation}
where $f_{S,\mu}(x)$ is the stable law of L\'evy index $\mu$. For $\mu =2$, the stable law is the Gaussian distribution  and one has $f_{S,2}(x+y)=1/(2\sqrt{\pi})\,\e^{-(x+y)^2/4}$. The right-hand side of\ (\ref{cross_4}) can then be computed explicitly, yielding ${\cal G}_2(u)$ as given in Eq.\ (\ref{mean_max_scaling.2}) with ${\cal G}_2(u) \sim 2/\sqrt{\pi}$ for $u \to 0$ and ${\cal G}_2(u) \sim 1/u$ as $u \to \infty$. For $1<\mu <2$, there is no explicit expression of the scaling function but the small and large argument behaviors can still be obtained. In the small $u$ limit, one has
\begin{equation}\label{cross_6}
{\cal G}_\mu(u)\sim\frac{\mu}{\pi}\, \Gamma\left(1-\frac{1}{\mu}\right)\ \ \ \ \ (u\to 0),
\end{equation}
while in the opposite large $u$ limit, one gets
\begin{equation}\label{cross_7}
{\cal G}_\mu(u)\sim
\frac{\Gamma(\mu -1)}{\pi(2-\mu)}\sin\left(\frac{\pi\mu}{2}\right)\, \frac{1}{u^{\mu -1}}
\ \ \ \ \ (u\to +\infty).
\end{equation}
From these small and large argument behaviors of ${\cal G}_\mu(u)$ one can check that the scaling form\ (\ref{cross_3}) reduces to the leading term of\ (\ref{eqRWEM0}) for $c=0$, and to Eq.\ (\ref{eqRWEMcMu}) for $1<\mu <2$ and $c\ne 0$, or Eq.\ (\ref{eqRWEMc2}) [with $\kappa_c$ given in Eq.\ (\ref{kappac})] for $\mu =2$ and a small nonzero $c$.

We now have all what is needed to derive the relations\ (\ref{eqFlux}) and\ (\ref{eqSurviv}) and obtain the large time asymptotic behavior of $\Phi_c(t)$ and $S_c(t)$.
%
%
\section{Smoluchowski flux problem in one dimension}\label{sec3}
\subsection{Continuous flux to a trap (Brownian motion)}\label{sec3.1}
\begin{figure}[t]
\includegraphics[width = \linewidth]{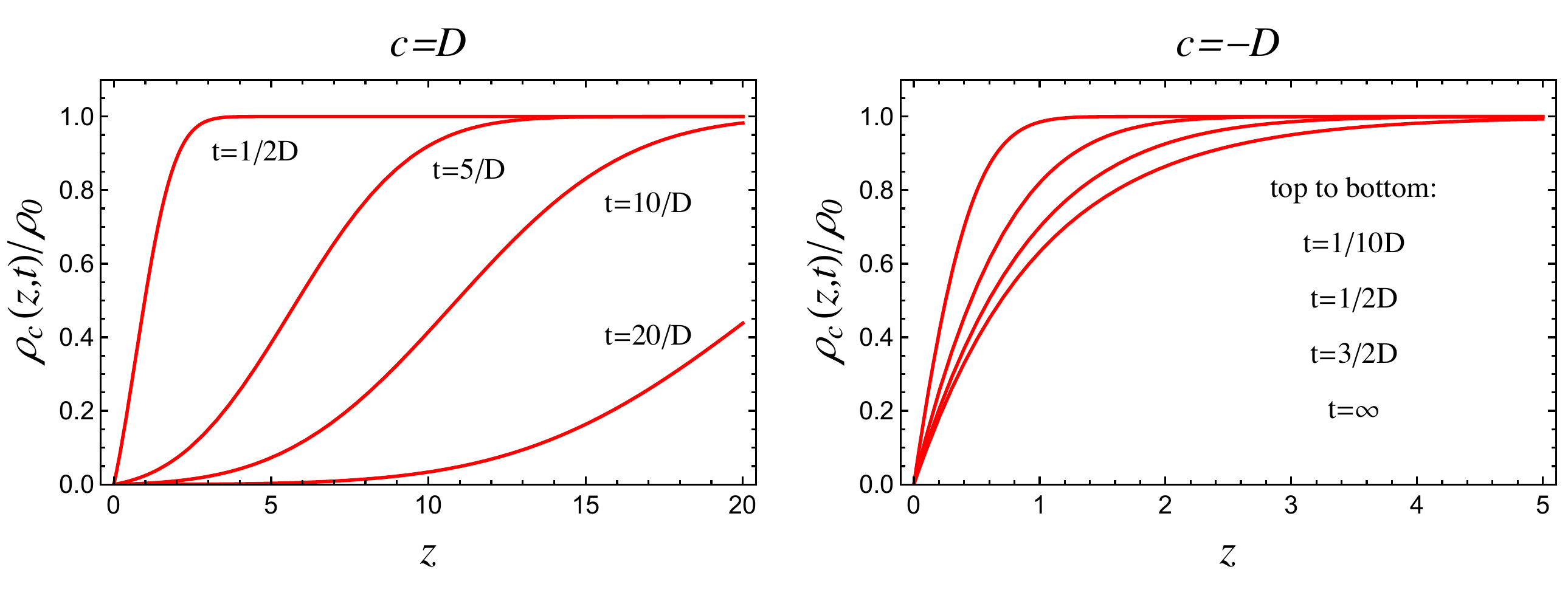}
\caption{Plots of the particle density profile $\rho_c(z,t)$ given in Eqs.\ (\ref{rho_Flux}) and\ (\ref{sol_max.1}) for $c=\pm D$ as a function of $z$ at different $t$. For $c>0$ (left), there is no limiting density profile and, at any fixed $z$, $\rho_c(z,t)$ goes to zero as $t\to\infty$. For $c<0$ (right), $\rho_c(z,t)$ converges rapidly to the limiting density profile $\rho_{c<0}(z,\infty)/\rho_0=1-\exp(-|c|z/D)$ (see Eq.\ (\ref{stat_dist.1})).} \label{Fig_rho}
\end{figure}

Consider a semi-infinite line $z\in\lbrack 0,\infty)$ with a single immobile trap, or absorber, at the origin $z=0$. Initially, the full semi-infinite line to the right of the trap is filled uniformly by non-interacting particles with constant density $\rho_0$. For $t>0$, the particles perform i.i.d. Brownian motions with a constant drift $c$, like in Eq.\ (\ref{eqBM}) (with a different starting position at $t=0$ for each particle). When a particle hits the trap at the origin, it gets absorbed there. Let $\rho_c(z,t)$, for $z\ge 0$, denote the density profile at time $t$, starting from a flat profile $\rho_c(z,0)=\rho_0$ (for $z>0$) at $t=0$. How does the density profile $\rho_c(z,t)$ evolve with time $t$? This is the $1D$ version of the celebrated Smoluchowski problem\ \cite{Smoluchowski}. Smoluchowski was also interested in the net flux $\Phi_c(t)$ out of the system up to time $t$\ \cite{Smoluchowski,Redner,Zif91,MCZ2006,ZMC2007}. Clearly,
\begin{equation}\label{int_Flux}
\Phi_c(t)=\int_{0}^{+\infty}\left\lbrack\rho_0 -\rho_c(z,t)\right\rbrack\, dz,
\end{equation}
counts the net flux out of the system up to time $t$ through {\it both boundaries} at $z=0$ and $z=\infty$. (Note that at the formal boundary $z=\infty$, particles can go both in and out of the system, unlike the boundary at $z=0$ where they only go out). The density profile $\rho_c(z,t)$ evolves via the biased diffusion equation\ \cite{Redner}
\begin{equation}\label{diff_rho}
\frac{\partial\rho_c(z,t)}{\partial t}=
D\frac{\partial^2\rho_c(z,t)}{\partial z^2}-c\frac{\partial\rho_c(z,t)}{\partial z},
\end{equation}
with boundary conditions $\rho_c(z=0,t)=0$ and $\rho_c(z\to\infty ,t)=\rho_0$, starting from the initial condition $\rho_c(z,t=0)=\rho_0$ for all $z>0$. By comparing this evolution equation to Eq.\ (\ref{bfp_max.1}) and the associated boundary and initial conditions\ (\ref{bc.1}) and\ (\ref{ic.1}), it follows immediately that
\begin{equation}\label{rho_Flux}
\rho_c(z,t)=\rho_0\, Q_c(z,t),
\end{equation}
where $Q_c(z,t)$ is given in Eq.\ (\ref{sol_max.1}). Figure\ \ref{Fig_rho} shows plots of $\rho_c(z,t)$ in Eq.\ (\ref{rho_Flux}) as a function of $z$ at different $t$ for both positive and negative $c$.
From this equation\ (\ref{rho_Flux}), we see that the Brownian particle density $\rho_c(z,t)$ in the Smoluchowski problem is somewhat unexpectedly related to the cumulative distribution of the maximum of a single Brownian motion starting at the origin. Using then Eq.\ (\ref{rho_Flux}) on the right-hand side of Eq.\ (\ref{int_Flux}), one obtains
\begin{equation}\label{BM_Flux_EM}
\Phi_c(t)=\rho_0\int_{0}^{+\infty}\left\lbrack 1-Q_c(z,t)\right\rbrack\, dz=\rho_0\, \E[M_c(t)],
\end{equation}
where we have used the expression\ (\ref{mean_max.1}) of $\E[M_c(t)]$. Thus, the net flux $\Phi_c(t)$ coincides with the expected maximum $\E[M_c(t)]$ up to a constant factor $\rho_0$, which proves the relation\ (\ref{eqFlux}).

From Eqs.\ (\ref{mean_max_scaling.1}) and\ (\ref{BM_Flux_EM}) one gets an exact expression of $\Phi_c(t)$ valid for all $c$ and $t$,
\begin{equation}
\Phi_c(t) = \rho_0\theta(c)\, ct + \rho_0\sqrt{D\, t}\, {\cal G}_2\left(\frac{|c|\sqrt{t}}{\sqrt{D}}\right).
\label{BM_Flux_scaling}
\end{equation}
In particular, using the behaviors of $\E[M_c(t)]$ in Eqs.\ (\ref{eqBMEM0bis}) and\ (\ref{eqBMEMcbis}), one obtains
\begin{equation}\label{BM_Flux_large_t}
\Phi_c(t)\sim\left\lbrace
\begin{array}{ll}
\rho_0 D/\vert c\vert,&c<0 \\
2\rho_0\sqrt{Dt/\pi},&c=0 \\
\rho_0 ct,&c>0
\end{array}\right.\ \ \ \ \ (t\to\infty).
\end{equation}
At first sight, in may seem a bit confusing that for $c>0$ the net flux out of the system increases linearly with $t$. This is simply due to the fact that $\Phi_c(t)$ includes contributions from both boundaries at $z=0$ and $z=\infty$. Hence, the net flux out of the system is not only through the origin but also `through infinity'. The dominant linear growth of $\Phi_c(t)$ for $c>0$ comes from the contribution of the boundary at $z=\infty$ (i.e. the outgoing flux `through infinity'). To make it clearer, it can be useful to give another, more transparent, derivation of the flux by integrating over the instantaneous current. To this end we first rewrite Eq.\ (\ref{diff_rho}) as a continuity equation
\begin{equation}\label{modifSatya1}
\frac{\partial\rho_c(z,t)}{\partial t}+\frac{\partial j_c(z,t)}{\partial z}=0,
\end{equation}
where the instantaneous current through $z$ (to the right) is
\begin{equation}\label{modifSatya2}
j_c(z,t)=-D\frac{\partial\rho_c(z,t)}{\partial z}+c\rho_c(z,t).
\end{equation}
The flux out of the system through $z=0$ up to time $t$ is $-\int_0^t j_c(0,\tau)\, d\tau$ and the flux out of the system `through infinity' up to time $t$ is $\int_0^tj_c(z\to \infty ,\tau)\, d\tau$. Hence, the net flux out of the system up to time $t$ is simply given by
\begin{equation}\label{modifSatya3}
\Phi_c(t)=-\int_0^t j_c(0,\tau)\, d\tau +\int_0^tj_c(z\to \infty ,\tau)\, d\tau ,
\end{equation}
which is equivalent to Eq.\ (\ref{int_Flux}) with the advantage of making explicit the contributions from both boundaries at $z=0$ and $z=\infty$. Now, for the Brownian case considered in this section, we can evaluate the individual fluxes through $z=0$ and $z=\infty$ by using Eq.\ (\ref{rho_Flux}) and the explicit solution in Eq.\ (\ref{sol_max.1}). After some straightforward algebra one finds, for any $c$,
\begin{equation}\label{modifSatya4}
-\int_0^t j_c(0,\tau)\, d\tau = \rho_0\sqrt{D\, t}\, {\cal G}_2\left(\frac{c\sqrt{t}}{\sqrt{D}}\right),
\end{equation}
where ${\cal G}_2(u)$ is given in Eq.\ (\ref{mean_max_scaling.2}), and
\begin{equation}\label{modifSatya5}
\int_0^tj_c(z\to \infty ,\tau)\, d\tau = \rho_0 ct.
\end{equation}
Injecting\ (\ref{modifSatya4}) and\ (\ref{modifSatya5}) onto the right-hand side of\ (\ref{modifSatya3}) yields
\begin{equation}\label{modifSatya6}
\Phi_c(t) = \rho_0 ct + \rho_0\sqrt{D\, t}\, {\cal G}_2\left(\frac{c\sqrt{t}}{\sqrt{D}}\right),
\end{equation}
which is exactly the same as Eq.\ (\ref{BM_Flux_scaling}) (it can be checked for positive and negative $c$, using the relation ${\cal G}_2(-u)=u+{\cal G}_2(u)$ in the latter case). For $c>0$ and $t\to\infty$, the dominant linear growth of $\Phi_c(t)$ in the third line of Eq.\ (\ref{BM_Flux_large_t}) corresponds clearly to the outgoing flux `through infinity' in Eq.\ (\ref{modifSatya5}). The fact that $\Phi_c(t)$ includes contributions from both boundaries at $z=0$ and $z=\infty$ is important and should be borne in mind to understand results that may seem paradoxical at first sight. We will get back to this point in a more general setting in Sec.\ \ref{sec4.3}.
%
%
\subsection{Discrete flux to a trap (random walks and L\'evy flights)}\label{sec3.2}
We now consider the same problem as above except that the particles perform i.i.d. discrete-time Markov jump processes (instead of Brownian motions), in the presence of a constant drift $c$, each evolving via Eq.\ (\ref{eqRW}). When a particle makes a jump to the negative side $z\le 0$, it gets absorbed by the trap. Let $\rho_c(z,n)$ denote the density profile at step $n$, starting from $\rho_c(z,0)=\rho_0$. The discrete-time counterpart of Eq.\ (\ref{int_Flux}) reads
\begin{equation}\label{int_Flux_discr}
\Phi_c(n)=\int_{0}^{+\infty}\left\lbrack\rho_0 -\rho_c(z,n)\right\rbrack\, dz,
\end{equation}
where, as above, $\Phi_c(n)$ counts the net average number of particles that have left the system up to step $n$ (through both boundaries at $0$ and $\infty$). One can easily write down the following recursion relation,
\begin{equation}\label{recursion_rho}
\rho_c(z,n)=\int_0^{+\infty}\rho_c(z^\prime ,n-1)\, f(z^\prime -z+c)\, dz^\prime ,
\end{equation}
where we have used $f(\eta)=f(-\eta)$, and with initial condition $\rho_c(z,0)=\rho_0$. The equation~(\ref{recursion_rho}) is the discrete-time counterpart of the diffusion equation\ (\ref{diff_rho}) in the Brownian motion case. Comparing Eq.\ (\ref{recursion_rho}) and Eq.\ (\ref{binteq_max}), and their associated initial conditions, one obtains
\begin{equation}\label{rho_Flux_discr}
\rho_c(z,n)=\rho_0\, Q_c(z,n),
\end{equation}
which, together with Eq.\ (\ref{int_Flux_discr}) and Eq.\ (\ref{mean_max.2}), yields
\begin{equation}\label{RW_Flux_EM}
\Phi_c(n)=\rho_0\int_{0}^{+\infty}\left\lbrack 1-Q_c(z,n)\right\rbrack\, dz=\rho_0\, \E[M_c(n)].
\end{equation}
This result enlarges the validity of the relation\ (\ref{eqFlux}) from continuous-time Brownian motion to discrete-time random walks.

Now, from the equation\ (\ref{RW_Flux_EM}) and the large $n$ expressions of $\E[M_c(n)]$ given in Sec.\ \ref{sec2.2} one can easily obtain the large $n$ behavior of $\Phi_c(n)$. In particular, the leading behavior of $\E[M_c(n)]$ in Eq.\ (\ref{cross_3}) gives the leading behavior of $\Phi_c(n)$ as
\begin{equation}\label{RW_Flux_scaling}
\Phi_c(n)\sim\rho_0\theta(c)\, cn +\rho_0an^{1/\mu} {\cal G}_\mu\left(\frac{|c|}{a} n^{1-1/\mu} \right) \:,
\end{equation}
and, from the large and small argument behaviors of the scaling function ${\cal G}_\mu (u)$, one has
\begin{equation}\label{RW_Flux_large_n2}
\Phi_c(n)\sim\left\lbrace
\begin{array}{ll}
\rho_0 a^2/\vert c\vert,&c<0 \\
2\rho_0 a\sqrt{n/\pi},&c=0 \\
\rho_0 cn,&c>0
\end{array}\right.\ \ \ \ \ (n\to\infty),
\end{equation}
for $\mu =2$, and
\begin{equation}\label{RW_Flux_large_nMu}
\Phi_c(n)\sim\left\lbrace
\begin{array}{ll}
\vert c\vert\rho_0 C\, n^{2-\mu}/(2-\mu),&c<0 \\
a\rho_0 \mu\Gamma(1-1/\mu)\, n^{1/\mu}/\pi,&c=0 \\
\rho_0 cn,&c>0
\end{array}\right.\ \ \ \ \ (n\to\infty),
\end{equation}
for $1<\mu <2$, where the constant $C$ is given in Eq.\ (\ref{const1}).
%
%
\section{Survival probability of the trap in one dimension: the lamb-lion problem}\label{sec4}
An other interesting question coming up in the setting of the Smoluchowski problem is the determination of the survival probability of the trap, $S_c(t)$, up to time $t$ (with either $t\in\mathbb{R}^+$ or $t=n\in\mathbb{N}$). More precisely, $S_c(t)$ is the probability that none of the particles, initially uniformly distributed with density $\rho_0$, has hit the origin up to time $t$. This problem is sometimes referred to in the literature as the lamb-lion problem\ \cite{Redner}: a lamb is immobile at the origin $z=0$, while $N$ lions with initial positions uniformly distributed over the interval $z\in[0,L]$ undergo independent Brownian motions (or random walks) with a constant drift~$c$. In the thermodynamic limit $L,\, N\to +\infty$ keeping $\rho_0 =N/L$ fixed, $S_c(t)$ is the survival probability of the lamb up to time $t$.

Let $S_c(t\vert N,L)$ denote the survival probability of the lamb up to time $t$ for given $N$ and $L$, and write $q_c(z_i ,t)$ the probability that the $i$-th lion, starting initially at $z_i\in[0,L]$, does not reach the lamb up to time $t$. Since the lions move independently from each other, one clearly has
\begin{eqnarray}\label{Surviv_NL}
S_c(t\vert N,L)&=&\left\langle\prod_{i=1}^N q_c(z_i ,t)\right\rangle
=\prod_{i=1}^N \left[\frac{1}{L}\int_0^L q_c(z_i ,t)\, dz_i\right] \\
&=& \left[\frac{1}{L}\int_0^L q_c(z ,t)\, dz\right]^N
= \left[1-\frac{1}{L}\int_0^L (1-q_c(z ,t))\, dz\right]^N \nonumber
\end{eqnarray}
where $\langle\cdot\rangle$ denotes the average over the $z_i$'s uniformly and independently distributed in $[0,L]$. Taking then the thermodynamic limit, $L,\, N\to +\infty$ with fixed $\rho_0 =N/L$, one gets
\begin{equation}\label{Surviv_th_limit}
S_c(t)=\lim_{L\to +\infty}S_c(t\vert N=\rho_0 \,L,L)=\exp\left[-\rho_0\int_0^{+\infty}(1-q_c(z,t))\, dz\right] .
\end{equation}
It remains to compute $q_c(z,t)$, which is done differently depending on whether the lions perform continuous-time Brownian motions or discrete-time random walks (or L\'evy flights).
%
%
\subsection{Continuous-time Brownian motions}\label{sec4.1}
In the case of Brownian motions, it is well known that $q_c(z,t)$ satisfies the backward Fokker-Planck equation\ \cite{pers_review}
\begin{equation}
\frac{\partial q_c(z,t)}{\partial t}= D\, \frac{\partial^2 q_c(z,t)}{\partial 
z^2} + c\, \frac{\partial q_c(z,t)}{\partial z}  \, ,
\label{bfp_Surviv_qc}
\end{equation}
valid for $z\ge 0$ with the boundary and initial conditions $q_c(0,t)=0$, $q_c(z\to\infty ,t)=1$, and $q_c(z,0)=1$ for all $z>0$. This is exactly the same equation as Eq.\ (\ref{bfp_max.1}), with the same boundary and initial conditions, in which $c$ is replaced with $-c$. It follows immediately that
\begin{equation}\label{BM_qc}
q_c(z,t)=Q_{-c}(z,t),
\end{equation}
with $Q_c(z,t)$ given in Eq.\ (\ref{sol_max.1}). Injecting\ (\ref{BM_qc}) into\ (\ref{Surviv_th_limit}) and using the equation\ (\ref{BM_Flux_EM}), one finds
\begin{equation}\label{BM_Surviv_EM}
S_c(t)=\exp\left[ -\Phi_{-c}(t)\right]
=\exp\left[ -\rho_0\, \E[M_{-c}(t)]\right] \;,
\end{equation}
which proves the relation\ (\ref{eqSurviv}) in the Brownian motion setting. From Eqs.\ (\ref{mean_max_scaling.1}) and\ (\ref{BM_Surviv_EM}) one gets an exact expression of $S_c(t)$ valid for all $c$ and $t$,
\begin{equation}\label{BM_Surviv_scaling}
S_c(t)=\exp\left[ \rho_0\theta(-c)\, ct
-\rho_0\sqrt{D\, t}\, {\cal G}_2\left(\frac{|c|\sqrt{t}}{\sqrt{D}}\right)\right] ,
\end{equation}
which, in the large time limit, reduces to
\begin{equation}\label{BM_Surviv_large_t}
S_c(t)\sim\left\lbrace
\begin{array}{ll}
K_{BM}\exp\left(-\rho_0 \vert c\vert t\right),&c<0 \\
\exp\left(-\frac{2\rho_0}{\sqrt{\pi}}\sqrt{Dt}\right),&c=0 \\
K_{BM},&c>0
\end{array}\right.\ \ \ \ \ (t\to\infty),
\end{equation}
where $K_{BM}=\exp(-D\rho_0/\vert c\vert)$. It follows in particular that if $c>0$ then the lamb will survive with a finite probability as $t\to\infty$.
%
%
\subsection{Generalisation to discrete-time random walks and L\'evy flights}\label{sec4.2}
In the case of discrete-time random walks evolving via Eq.\ (\ref{eqRW}), the time evolution of $q_c(z,n)$ is given by the recursion relation\ \cite{pers_review}
\begin{equation}\label{binteq_Surviv}
q_c(z,n)=\int_0^{+\infty}q_c(z^\prime ,n-1)\, f(z^\prime -z-c)\, dz^\prime ,
\end{equation}
with the initial condition $q_c(z,0)=1$ for all $z\ge 0$. This integral equation is the discrete-time counterpart of the continuous-time backward Fokker-Planck equation\ (\ref{bfp_Surviv_qc}). It is readily seen that the equation\ (\ref{binteq_Surviv}) for $q_c(z,n)$ is exactly the same as the one for $Q_{-c}(z,n)$, Eq.\ (\ref{binteq_max}) with the same initial condition and $c$ replaced with $-c$. It thus follows immediately that
\begin{equation}\label{RW_qc}
q_c(z,n)=Q_{-c}(z,n).
\end{equation}
Finally, injecting\ (\ref{RW_qc}) into\ (\ref{Surviv_th_limit}), with $t=n\in\mathbb{N}$, and using the equation\ (\ref{RW_Flux_EM}), one obtains
\begin{equation}\label{RW_Surviv_EM}
S_c(n)=\exp\left[ -\Phi_{-c}(n)\right]
=\exp\left[ -\rho_0\, \E[M_{-c}(n)]\right] \;,
\end{equation}
which proves the relation\ (\ref{eqSurviv}) for discrete-time random walks.

In the large $n$ limit, the leading exponential behavior of $S_c(n)$ is correctly obtained by replacing $\E[M_{-c}(n)]$ with the scaling form\ (\ref{cross_3}), but it misses the slowly varying prefactor associated with the subdominant corrections to $\E[M_{-c}(n)]$, as these corrections do not appear in Eq.\ (\ref{cross_3}). Note that knowing this prefactor is important because it can contribute significantly to the order of magnitude of $S_c(n)$ when $n$ is large. For $\mu =2$ the prefactor is found to reduce to a mere constant and, according to Eqs.\ (\ref{RWasym0}) and\ (\ref{RWasymMu=2}), one has
\begin{equation}\label{RW2_Surviv_large_t}
S_c(n)\sim\left\lbrace
\begin{array}{ll}
K_{RW}\exp\left(-\rho_0 \vert c\vert n\right),&c<0 \\
K_{RW}^{(0)}\exp\left(-\frac{2a\rho_0}{\sqrt{\pi}}\sqrt{n}\right),&c=0 \\
K_{RW},&c>0
\end{array}\right.\ \ \ \ \ (n\to\infty),
\end{equation}
as announced in Eq.\ (\ref{eqRWSurviv2}), where $K_{RW}^{(0)}=\exp(-a\rho_0\gamma)$ with $\gamma$ given by Eq.\ (\ref{gamma}), and $K_{RW}=\exp(-\vert c\vert\rho_0\kappa_c)$ with $\kappa_c$ given by Eq.\ (\ref{kappac}) (note that $\kappa_{-c}=\kappa_c$). For $1<\mu <2$, the large $n$ behavior of $S_c(n)$ is obtained by using the equations\ (\ref{RWasym0}),\ (\ref{RWasymMu.1}), or\ (\ref{RWasymMu.2}) on the right-hand side of\ (\ref{RW_Surviv_EM}). Let $\mathscr{S}_{>}(n)$ and $\mathscr{S}_{<}(n)$ be defined respectively by
\begin{equation}\label{prefactor1}
\ln\mathscr{S}_{>}(n)=-\rho_0\vert c\vert
\sum_{m=2}^{[1/(\mu -1)]}\frac{C_m n^{1-m(\mu -1)}}{1-m(\mu -1)}
-\rho_0\vert c\vert\kappa_c ,
\end{equation}
if $\mu\ne 1+1/p$ for any integer $p$,
\begin{equation}\label{prefactor2}
\ln\mathscr{S}_{>}(n)=-\rho_0\vert c\vert
\sum_{m=2}^{p-1}\frac{p\,C_m n^{1-m/p}}{p-m}
-\rho_0\vert c\vert C_p\ln n
-\rho_0\vert c\vert\kappa_c ,
\end{equation}
if $\mu =1+1/p$ for some integer $p$, and
\begin{equation}\label{prefactor3}
\mathscr{S}_{<}(n)=\mathscr{S}_{>}(n)
\exp\left(-\frac{\vert c\vert\rho_0 C}{2-\mu}\, n^{2-\mu}\right).
\end{equation}
We recall that the constants $C$ and $C_m$ are given in Eqs. (\ref{const1}) and (\ref{constCm}) respectively.   
Bringing out the leading exponential behavior of $S_c(n)$, one finds
\begin{equation}\label{RWMu_Surviv_large_t}
S_c(n)\sim\left\lbrace
\begin{array}{ll}
\mathscr{S}_{<}(n)\exp\left(-\rho_0 \vert c\vert n\right),&c<0 \\
K_{RW}^{(0)}\exp\left(-\frac{a\rho_0\mu\Gamma(1-1/\mu)}{\pi}n^{1/\mu}\right),&c=0 \\
\mathscr{S}_{>}(n)\exp\left(-\frac{c\rho_0 C}{2-\mu}n^{2-\mu}\right),&c>0
\end{array}\right.\ \ \ \ \ (n\to\infty),
\end{equation}
as announced in Eq.\ (\ref{eqRWSurvivMu}), where the prefactors $\mathscr{S}_{<}(n)$ and $\mathscr{S}_{>}(n)$ are slowly varying compared with the exponential of $n$ and $n^{2-\mu}$, respectively [see Eqs. (\ref{prefactor3}) and (\ref{prefactor2})]. Note that the numerical value of the constant $K_{RW}^{(0)}$ in Eq.\ (\ref{RWMu_Surviv_large_t}) is not the same as in Eq.\ (\ref{RW2_Surviv_large_t}), as $\gamma$ in Eq.\ (\ref{gamma}) depends on $\mu$. As explained in the introduction, an  interesting, and a bit counterintuitive, consequence of this result is that the survival probability of the lamb still decays to zero in the presence of a positive drift, unlike in the $\mu =2$ case where it goes to a constant, as can be seen in the third line of Eq.\ (\ref{RWMu_Surviv_large_t}). Such a decrease of $S_c(n)$ for $c>0$ and $1<\mu <2$  is to be attributed to the fact that, sooner or later, lions undergoing L\'evy flights will always perform rare big jumps that will overcompensate for their linear drift away from the lamb.
%
%
\subsection{An apparent paradox}\label{sec4.3}
We end this section with the following interesting observation. The relation $S_c(t)=\exp[-\Phi_{-c}(t)]$ (with $t\in\mathbb{R}^+$ or $t=n\in\mathbb{N}$) indicates that the survival probability of the lamb depends on the {\it total} net flux of lions out of the system, i.e., through both boundaries at $z=0$ and $z=\infty$. On the other hand, intuitively one expects the survival probability to depend on the flux of lions through the origin only, where the immobile lamb is located. Although seemingly paradoxical, it can be shown that, actually, there is no contradiction whatsoever, as we will now see. Below, we will use the word `particles' for both particles in the Smoluchowski problem and lions in the lamb-lion problem. Before resolving the (apparent) contradiction, define $\phi_c(z_0,t) = -\int_0^t j_c(z_0,\tau)\, d\tau$ where $j_c(z_0,\tau)$ denotes the instantaneous current of particles through $z=z_0$ counted algebraically. More precisely, $\phi_c(z_0,t)$ is the average number of particles that have crossed $z=z_0$ from $z>z_0$ to $z<z_0$ minus the average number of particles that have crossed $z=z_0$ from $z<z_0$ to $z>z_0$, up to time $t$. Note that, since no particle enters the system at $z=0$ (from the negative side), $\phi_c(0,t)$ reduces to the average number of particles that have passed through the origin (to the negative side), up to time $t$.

First we prove that $S_c(t)$ does depend on $\phi_c(0,t)$ only, as expected. Clearly, the probability for a given particle starting at some $z>0$ to cross the origin before time $t$ is $1-q_c(z,t)$. Now, the average number of particles initially in $[z,\, z+dz]$ is $\rho_0 dz$ and $\rho_0 (1-q_c(z,t))\, dz$ gives the average number of particles initially in $[z,\, z+dz]$ that have crossed the origin before time $t$. Hence, integrating over the initial position of the particles gives the average number of particles that have crossed the origin up to time $t$,
\begin{equation}\label{fluxat0}
\phi_c(0,t)=\rho_0\int_0^{+\infty}\lbrack 1-q_c(z,t)\rbrack\, dz.
\end{equation}
Consequently, putting\ (\ref{fluxat0}) on the right-hand side of Eq.\ (\ref{Surviv_th_limit}), one obtains
\begin{equation}\label{new_eqSurviv}
S_c(t)=\exp\lbrack -\phi_c(0,t)\rbrack .
\end{equation}
This relation is the one we should expect intuitively. Indeed, the passage of particles through the origin forms a Poisson process of parameter $\phi_c(0,t)$ and the probability that no particle has passed through the origin up to time $t$ is just $\exp\lbrack -\phi_c(0,t)\rbrack$. Comparing the equation\ (\ref{new_eqSurviv}) with $S_c(t)=\exp[-\Phi_{-c}(t)]$ in Eq.\ (\ref{eqSurviv}) leads to a nontrivial relation,
\begin{equation}\label{flux_relation}
\Phi_{-c}(t)=\phi_c(0,t),
\end{equation}
which is thus necessary to resolve the above-mentioned apparent contradiction. Changing the sign of $c$ from one side of Eq.\ (\ref{flux_relation}) to the other is crucial. Clearly, $\Phi_c(t)\ne\phi_c(0,t)$, and we would have to face a real contradiction if there was no such a change of sign.

It remains to prove the identity\ (\ref{flux_relation}). To this end, we use the relation
\begin{equation}\label{symmetry1}
\E[M_{-c}(t)]=\E[M_c(t)]-ct,
\end{equation}
discussed at the end of the introduction in Ref.\ \cite{MMS2018} for $t=n\in\mathbb{N}$ and valid also for $t\in\mathbb{R}^+$. (The reasoning leading to\ (\ref{symmetry1}) in\ \cite{MMS2018} is quite general and it can be transposed straightforwardly to the case of continuous-time processes). From Eqs.\ (\ref{eqFlux}) and\ (\ref{symmetry1}) one gets
\begin{equation}\label{symmetry2}
\Phi_{-c}(t)=\Phi_c(t)-\rho_0 ct=\phi_c(0,t)-\phi_c(z\to\infty ,t)-\rho_0 ct,
\end{equation}
where we have written $\Phi_c(t)$ as $\Phi_c(t)=\phi_c(0,t)-\phi_c(z\to\infty ,t)$. Let us now estimate $\phi_c(z\to\infty ,t)$. This is precisely given by
\begin{eqnarray}\label{phic_zlarge}
\phi_c(z\to\infty ,t) = -\int_0^t j_c(z \to \infty, \tau) \, d\tau 
\end{eqnarray}
where $j_c(z \to \infty, \tau)$ is the instantaneous current at large $z$. Using the fact that the particles very far from the absorbing boundary do not feel the presence of the boundary, we see that, as $z\to\infty$, the particle density profile goes to the initial flat one with density $\rho_0$ and the average particle velocity goes to the drift $c$, from which it follows that  $j_c(z \to \infty,t) = c\,\rho_0$. Hence, integrating over time this instantaneous current in Eq. (\ref{phic_zlarge}), we get $\phi_c(z \to \infty,t) = - \rho_c \, ct$ which generalizes the Brownian motion result in Eq.\ (\ref{modifSatya5}). Putting this result on the right-hand side of Eq.\ (\ref{symmetry2}) yields $\Phi_{-c}(t)=\phi_c(0,t)$, which proves the relation~(\ref{flux_relation}).
%
%
\section{Numerical simulations}\label{sec5}
{We have compared our analytical results with numerical simulations. For this purpose, we have simulated $N$ random walks with a drift evolving via Eq. (\ref{eqRW}). Initially, the $N$ walkers are uniformly distributed over the interval $[0,L]$ with a uniform density $\rho_0$. To compare
with our results we have to consider the limit $N,L \to \infty$ with $\rho_0 = N/L$ fixed. In the simulations presented here, we have taken $N = L = 10^4$, corresponding to $\rho_0 = 1$. We have first checked our predictions for the Smoluchovski problem and computed the flux of particles $\Phi_c(n)$ out of the system. We have checked the exact identity $\Phi_c(n) = \rho_0 \, \mathbb{E}[M_c(n)]$. Hence our numerical results for $\Phi_c(n)$ reproduce the ones obtained previously for $\mathbb{E}[M_c(n)]$ in Ref. \cite{MMS2018} exactly. Therefore, we do not not show them again here and referred the interested reader to the section 6 of Ref. \cite{MMS2018}. Instead, in this section we present our numerical results for the survival probability $S_c(n)$. 

\begin{figure}[t]
\includegraphics[width = \linewidth]{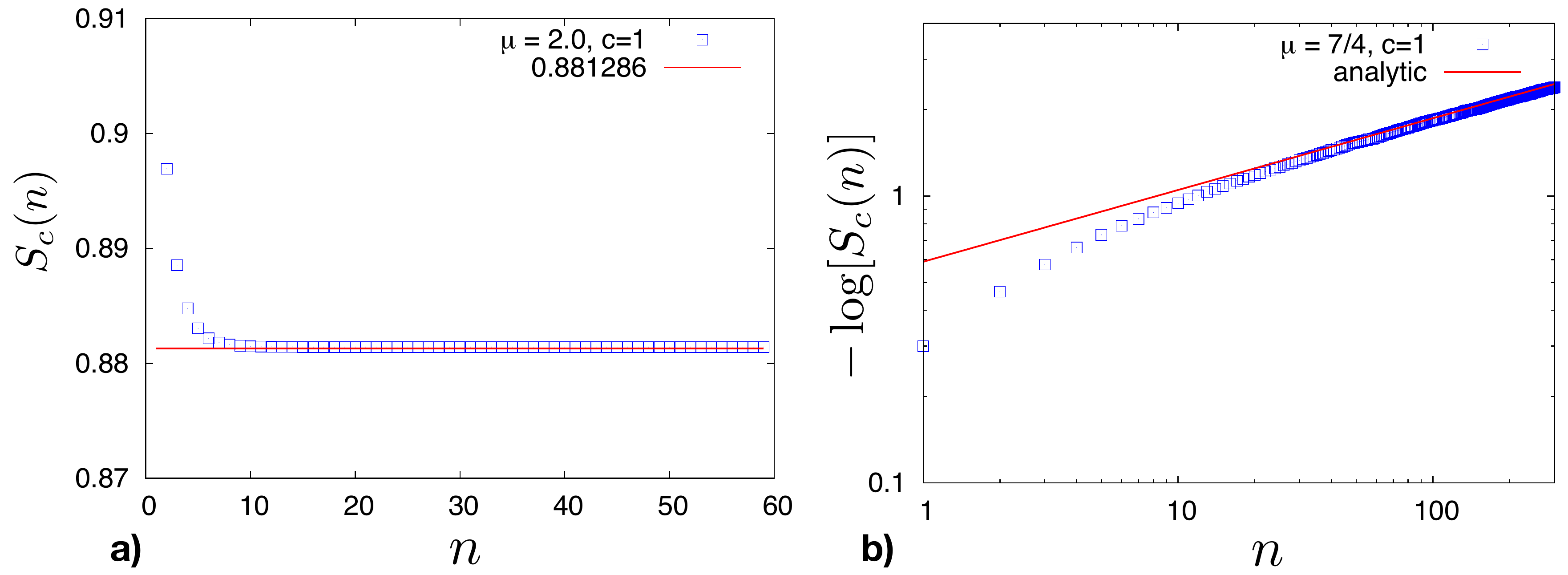}
\caption{{\bf a)} Plot of the survival probability $S_c(n)$ as a function of $n$ for particles/lions performing a Gaussian random walk with unit variance $\sigma=1$ (hence $\mu = 2$) and with a drift $c=1$. The blue squares show the results obtained by simulating $N=10^4$ random walks with a uniform density $\rho_0=1$. The solid line corresponds to our theoretical prediction given in the third line of Eq. (\ref{eqRWSurviv2}), i.e. $S_c(n \to \infty) \to K_{RW} = 0.881286\ldots$. {\bf b)} Plot of $-\log[S_c(n)]$ as a function of $n$, on a log-log plot, for L\'evy flights with $\mu = 7/4$ and drift $c=1$. Here also the blue squares show the results obtained by simulating $N=10^4$ random walks with a uniform density $\rho_0=1$ while the solid line corresponds to the leading behavior as predicted by the third line of Eq. (\ref{eqRWSurvivMu}), i.e. $-\log[S_c(n)] \approx c\rho_0 \,C/(2-\mu) n^{2-\mu}$ with $C$ given in (\ref{const1}). A comparison of these two plots {\bf a)} and {\bf b)} clearly shows that $S_c(t)$ is much smaller for L\'evy flights with $\mu <2$ than for standard random walks (with $\mu=2$).} \label{Fig_S_cplus}
\end{figure}
\noindent{\it The case $c>0$}. In the left panel of Fig. \ref{Fig_S_cplus} we show a plot of our numerical data obtained for $S_c(n)$ for the Gaussian random walk, i.e. for a random walk (\ref{eqRW}) with a Gaussian jump distribution $f(\eta)$ with variance $\sigma = 1$, thus corresponding to a L\'evy index $\mu=2$, and with a drift $c=1$. In this case, the constant $\kappa_c$ entering the definition of the amplitude $K_{RW} = \exp(-|c| \rho_0 \kappa_c)$ in Eq. (\ref{eqRWSurviv2}) is given by the expression \cite{MMS2018}
\begin{eqnarray}\label{kappac_Gauss}
\kappa_c = \sum_{m=1}^\infty \left[\frac{\e^{-b^2 m}}{2b\sqrt{\pi m}} - \frac{1}{2} {\rm erfc}(b\sqrt{m}) \right]
\end{eqnarray}
where $b = |c|/\sigma\sqrt{2}$ and ${\rm erfc}(z) = 2/\sqrt{\pi} \int_x^\infty \e^{-t^2}\, dt$. On the left panel of Fig. \ref{Fig_S_cplus}, we see that $S_c(n \to \infty) \approx K_{RW}$ for $n\gtrsim 10$ steps, in agreement with our predictions in the third line of Eq. (\ref{eqRWSurviv2}). In addition, the numerical value of $K_{RW}$ is also found to be in very good agreement with the theoretical one: for $c=1$ and $\sigma = 1$, one has $\kappa_c = 0.126373 \ldots$, which gives $K_{RW} = 0.881286 \ldots$ for $\rho_0 = 1$. In the right panel of Fig. \ref{Fig_S_cplus} we show a plot of $-\log [S_c(n)]$ as a function of $n$ for a L\'evy flight of index $\mu = 7/4$. From our analytical predictions in the third line of Eq. (\ref{eqRWSurvivMu}) one expects that, for large $n$, $-\log [S_c(n)] = c \rho_0 C/(2-\mu) \, n^{2-\mu} + o(n^{2-\mu})$ where the constant $C$ is given in Eq. (\ref{const1}). This leading behavior is indicated as a solid line in Fig. \ref{Fig_S_cplus} and we see that the agreement with our numerical data is quite good for $n \gtrsim 100$.  
\begin{figure}[t]
\includegraphics[width = \linewidth]{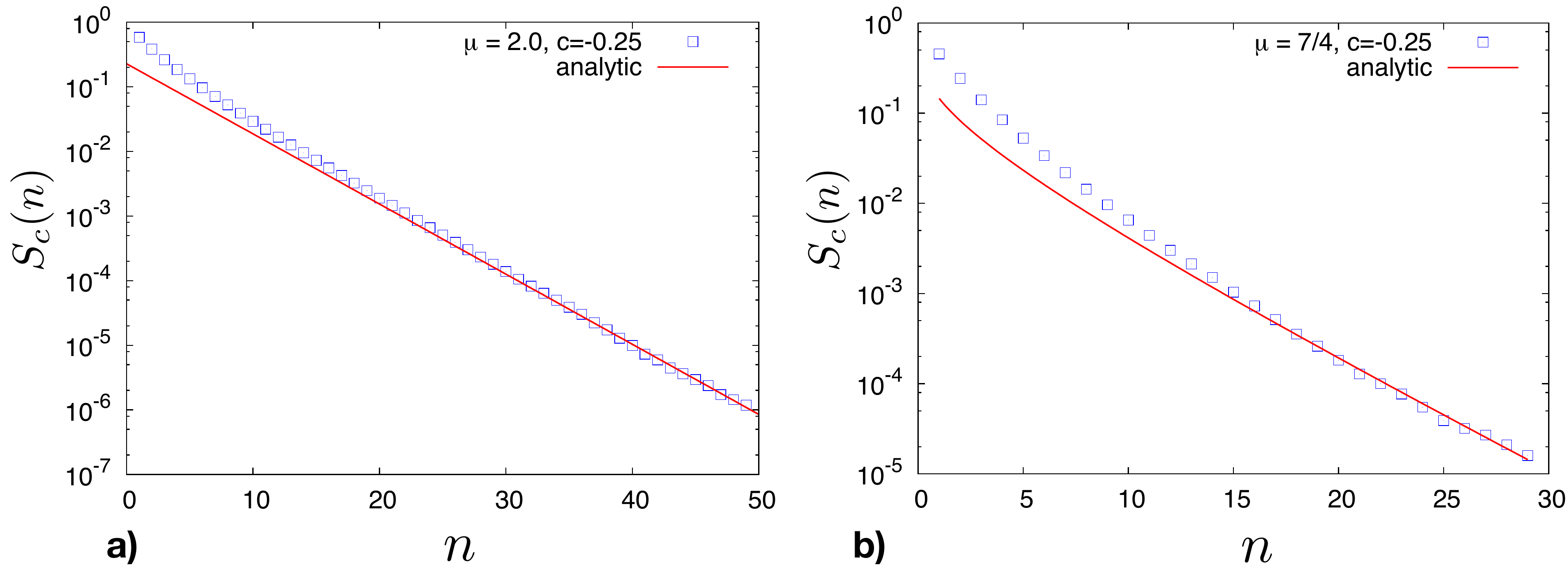}
\caption{{\bf a)} Plot of the survival probability $S_c(n)$ as a function of $n$ (on a semi-log scale) for particles/lions performing a Gaussian random walk with unit variance $\sigma=1$ (hence $\mu = 2$) and with a drift $c=-1/4$. The blue squares show the results obtained by simulating $N=10^4$ random walks with a uniform density $\rho_0=1$. The solid line corresponds to our theoretical prediction given in the first line of Eq.~(\ref{eqRWSurviv2}), i.e. $S_c(n) \approx K_{RW} \exp(-\rho_0\,|c| n)$ with $K_{RW} = 0.228255 \ldots$. {\bf b)} Plot of $S_c(n)$ as a function of $n$ (on a semi-log scale) for L\'evy flights with $\mu = 7/4$ and drift $c=-1/4$. Here also the blue squares show the results obtained by simulating $N=10^4$ random walks with a uniform density $\rho_0=1$ while the solid line corresponds to the leading behavior as predicted by the first line of Eq. (\ref{eqRWSurvivMu}), i.e. $S_c(n) \approx \exp{\left(- |c| \rho_0 n - |c|\rho_0 C/(2-\mu) n^{2-\mu}\right)}$ with $C$ given in (\ref{const1}).} \label{Fig_S_cminus}
\end{figure}
\noindent{\it The case $c<0$}. In the left panel of Fig. \ref{Fig_S_cminus} we show a plot of our numerical data obtained for $S_c(n)$ for the same Gaussian random walk as in the case of the left panel of Fig. \ref{Fig_S_cplus} ($\mu =2$) but now with a negative drift $c=-1/4$. Comparing these data with our theoretical prediction given in the first line of Eq. (\ref{eqRWSurviv2}) with $K_{RW} = 0.228255 \ldots$, we find that the agreement is very good for $n\gtrsim 15$, as can be seen on the left panel of Fig. \ref{Fig_S_cminus}. Finally, in the right panel of Fig. \ref{Fig_S_cminus} we compare the results of our numerical simulations for L\'evy flights with index $\mu = 7/4$ and drift $c=-1/4$ with our theoretical predictions given in the first line of Eq. (\ref{eqRWSurvivMu}), i.e. in this case where $\mu = 7/4$, $S_c(n) = \exp{\left(- |c| \rho_0 n - |c|\rho_0 C/(2-\mu) n^{2-\mu} + O(1)\right)}$ for $n \to \infty$. Again the agreement is very good for $n\gtrsim 15$. 
}

%
\section{Conclusion}\label{sec6}
{In this paper, we have generalized the Smoluchowski flux and the lamb-lion problems, well studied for Brownian motion, to discrete-time random walks/L\'evy flights with a constant drift. Specifically, we have considered the problem of independent particles performing one-dimensional random walks or L\'evy flights, with L\'evy index $1< \mu \leq 2$ and a constant drift $c$, in the presence of an absorbing immobile trap (or lamb) at the origin. Initially these particles are uniformly distributed over the positive real axis with density $\rho_0$. We have obtained exact results for the net flux $\Phi_c(n)$ of particles out of the system (Smoluchowski problem) and for the survival probability $S_c(n)$ of the trap or lamb after $n$ steps (lamb-lion problem). After deriving the important relations $\Phi_c(n)=\rho_0\, \E[M_c(n)]$ and $S_c(n)=\exp[ -\Phi_{-c}(n)]$ linking $\Phi_c(n)$ and $S_c(n)$ to the expected value ${\mathbb{E}}[M_c(n)]$ of the maximum of a single random walker in the presence of a drift $c$, we have used  our recently obtained exact results for ${\mathbb{E}}[M_c(n)]$ in the large $n$ limit \cite{MMS2018}. We have found in particular that, for lions undergoing independent L\'evy flights ($1<\mu <2$), the survival probability of the lamb still decays to zero even in the presence of a positive drift  ($c>0$). More precisely, one has $S_{c>0}(n \to \infty) \approx \exp\left(-\lambda \, n^{2-\mu}\right)$ for $1< \mu < 2$, where $\lambda$ is a $\mu$-dependent positive constant, while $S_{c>0}(n \to \infty) \to K_{RW} > 0$ for standard random walks (i.e., with $\mu = 2$). This somewhat counterintuitive result follows from the fact that lions undergoing L\'evy flights will always perform rare big jumps that will overcompensate for their linear drift away from the lamb. We have checked that our analytical results are confirmed by numerical simulations.

The present work provides an example of interesting physical applications of extreme value statistics of random walks and L\'evy flights. For discrete time random walks/L\'evy flights, this extreme value statistics problem is surprisingly very hard, though several exact analytical results have been obtained recently~\cite{MMS2018,MMS13,MMS14}. It would be interesting to find other applications of these results. For instance, in the absence of a drift (i.e., $c=0$), the statistical properties of the convex hull a two-dimensional random walk (or Brownian motion) can be computed from the extreme value statistics of the corresponding one-dimensional random walk (or Brownian motion) \cite{GLM2017,RMC2009,MRC2010,DMRZ13}. The next step could be to extend these results and study the convex hull of random walks or Brownian motion in the presence of a drift $c\neq 0$. 

An other natural continuation of this work would be to extend the results of this paper to other stochastic processes like, e.g., persistent random walks, i.e. particles performing ``run and tumble'' dynamics which are currently widely studied in the context of active matter (see e.g. \cite{Be2014, TC2008}). Recently, the dynamics of such run and tumble particles in semi-infinite domains have been worked out \cite{Malakar,EM2018,DMS2019,MW1993,SK2019} and it would be interesting to extend the present studies to the case of these active particles.
}
%
%

%
%
\end{document}